\documentstyle[aps,preprint]{revtex}
\begin{document}

\title{A model for the size distribution of customer groups and businesses}
\author{Dafang Zheng$^{1,2,}$\thanks
{e-mail: phdzheng@scut.edu.cn},
G. J. Rodgers$^{1}$ and P. M. Hui$^{3}$}
\address{$^{1}$ Department of Mathematical Sciences, Brunel  University,
 Uxbridge, Middlesex UB8 3PH, UK}
\address{$^{2}$ Department of Applied Physics, South China University of
Technology, \\
Guangzhou 510641, P.R. China}
\address{$^{3}$ Department of Physics, The Chinese University of Hong
Kong, \\
Shatin, New Territories, Hong Kong}
\maketitle

\begin{abstract}
We present a generalization of the dynamical model of information
transmission and herd behavior proposed by Egu\'{\i}luz and
Zimmermann.  A characteristic size of group of agents $s_{0}$ is
introduced.  The fragmentation and coagulation rates of groups of
agents are assumed to depend on the size of the group. We present
results of numerical simulations and mean field analysis. It is
found that the size distribution of groups of agents $n_{s}$
exhibits two distinct scaling behavior depending on $s \leq s_{0}$
or $s > s_{0}$.  For $s \leq s_{0}$, $n_{s} \sim s^{-(5/2 +
\delta)}$, while for $s > s_{0}$, $n_{s} \sim s^{-(5/2 -\delta)}$,
where $\delta$ is a model parameter representing the sensitivity
of the fragmentation and coagulation rates to the size of the
group. Our model thus gives a tunable exponent for the size
distribution together with two scaling regimes separated by a
characteristic size $s_{0}$.  Suitably interpreted, our model can
be used to represent the formation of groups of customers for
certain products produced by manufacturers.  This, in turn, leads
to a distribution in the size of businesses.  The characteristic
size $s_{0}$, in this context, represents the size of a business
for which the customer group becomes too large to be kept happy
but too small for the business to become a brand name.  

\end{abstract}

\newpage
\section{Introduction}
Power law behavior of various kinds have been observed in a wide
range of systems.  Within the context of economics and finance, it
has been found, for example, that the wealth of
individuals\cite{pareto}, price-returns in stock
markets\cite{stanley1,stanley2}, and company sizes in different
countries all give non-trivial and interesting
distributions\cite{stanley3,nagel,ramsden}.  By analyzing U.S.
establishment and firm sizes, Nagel {\em et al.}\cite{nagel} found
that the size distributions $n_{s}$ are power laws of the form
$n_{s} \sim s^{-\tau}$ with $\tau=2$ for firms with annual sales
larger than some typical size.  Income distribution of companies
and size distribution of debts among bankrupt companies in
Japan\cite{okuyama,aoyama} also indicate a power law behavior with
an exponent of about $-2$.  However, studies on the size
distribution of companies in different
countries\cite{ramsden,takayasu} indicated that the distributions
as well as the exponents are different from country to country and
hence are {\em not} universal.

Nagel {\em et al.} introduced several models of price formation
using a two-dimensional spatial structure for information
transmission in which consumers can only learn from their
neighbors\cite{nagel}. The formation of groups of consumers with
shared opinion has direct impacts on the price-returns in stock
markets\cite{cont,EZ} and on the size of
businesses\cite{nagel,rodgers1} as consumers of similar opinion
may act collectively in making a transaction or a deal.  Such herd
formation and information transmission in a population have been a
topic of active research in recent years.  Recently, Egu\'{\i}luz
and Zimmermann\cite{EZ} proposed and studied a simple model
(henceforth referred to as the EZ model) of stochastic opinion
cluster formation and information dispersal.  It is a dynamical
model in which there is a continual grouping and re-grouping of
agents to form clusters of similar opinion.  A group of agents act
together in a transaction and the group dissolves after the
transaction has occurred.  When a group decides not to trade, it
may combine with another group to form a larger group. Detailed
numerical studies\cite{EZ} and mean field analysis\cite{rodgers2}
revealed that the model could lead to a fat-tail distribution of
price returns qualitatively similar to those observed in real
markets\cite{stanley1,stanley2}.  The size distribution $n_{s}$ of
groups is shown to take on the form $n_{s} \sim s^{-5/2}$ for a
range of group size $s$, followed by an exponential
cutoff\cite{EZ,rodgers2}. The model represents a dynamic
generalization of the percolation-type model previously introduced
by Cont and Bouchard\cite{cont} in which the behavior $n_{s} \sim
s^{-5/2}$ is also observed.  Interesting, it was recently found
that the exponent characterizing this scaling behavior can
actually be changed and made tunable by introducing a
size-dependent dissociation and coagulation rates of
groups\cite{zheng1} in the EZ model. The possibility of tuning the
exponent makes the EZ model and its generalizations a good
starting point for modelling herd formation and opinion sharing in
different contexts.

In this paper, we propose and study an extension of the EZ model
in which a characteristic size $s_{0}$ of group of agents
(customers) is introduced. The rates of dissociation and
coagulation of groups are assumed to take on different forms for
$s \leq s_{0}$ and for $s>s_{0}$.  The distribution in the size of
groups of agents is found to exhibit two scaling regimes. Within
the context of customers, the different behavior for $s \leq
s_{0}$ and $s > s_{0}$ represent the situations in which a small
group of customers may find themselves better served by businesses
and customers tend to be happy when they find themselves to be
part of a big group of customers. The distribution of groups of
customers is unevitably related to the distribution of sizes of
businesses that the customers support. In Sec.II, we introduce the
model and present results of numerical simulations.  Results of a
mean field analysis are given in Sec.III and compared with
numerical results. We summarize our results in Sec.IV.

\section{The model}

Our model is an extension of the EZ model\cite{EZ}, which, in
turn, is a dynamical version of the static percolation model of
Cont and Bouchaud\cite{cont}. Consider a system with a total of
$N$ agents (customers). These agents are organized into groups.
They share the same information and hence they act collectively,
i.e., customers belonging to a group decide whether to make a deal
and what to do after a deal collectively. Initially, all the
customers are isolated, i.e., each group has only one customer. At
each time step, a customer $i$ is selected at random. With
probability $a$, all the customers belonging to the group of
customer $i$ decide to make a deal, say, with a certain business
or to buy a certain product from a manufacturer. After the deal,
the group of customers will find themselves disappointed with
probability $f(s_{i})$.  The group is then broken up into isolated
customers and the corresponding business loses her customers.
With probability $(1-f(s_{i}))$, the group of customers are
pleased with the deal (service) and they remain in the group. With
probability $1-a$, the chosen customer $i$ and its group decide
not to make a deal.  In this case, another customer $j$ is
selected at random among the whole population of customers.  With
probability $c(s_{i}, s_{j})$, the two groups of customers $i$ and
$j$ coagulate to form a bigger group of customers, and with
probability $(1-c(s_{i}, s_{j}))$ they remain separate. We study
the size distribution of the groups of customers which is
unevitably related to the size distribution of the businesses that
the customers support.

In the present work, we take the fragmentation rate of group of customers
to be
\begin{equation}
    f(s_{i})=\left\{
            \begin{array}{ll}
                p(s_{i}), & s_{i} \leq s_{0}, \\
                q(s_{i}), & s_{i} > s_{0},\\
            \end{array}
        \right.
\label{eq: f}
\end{equation}
and the coagulation rate of groups of customers as
\begin{equation}
    c(s_{i}, s_{j})=\left\{
            \begin{array}{ll}
                p(s_{i})p(s_{j}), & s_{i} \leq s_{0}, s_{j} \leq s_{0},\\
                p(s_{i})q(s_{j}), & s_{i} \leq s_{0}, s_{j} > s_{0},\\
                q(s_{i})p(s_{j}), & s_{i} > s_{0}, s_{j} \leq s_{0},\\
                q(s_{i})q(s_{j}), & s_{i} > s_{0}, s_{j} > s_{0}.\\
            \end{array}
        \right.
\label{eq: c}
\end{equation}
The functions $p(s)$ and $q(s)$ are taken to be increasing and
decreasing power law functions:
\begin{equation}
p(s)=(\frac{s}{s_{0}})^{\delta}, \; \; q(s)=(\frac{s_{0}}{s})^{\delta},
\label{eq: pq}
\end{equation}
where a characteristic group size of customers $s_{0}$ is
introduced. From the standpoint of customers, a small group of
customers may find themselves easier to be pleased by the service
and the deal. Customers are also happier when they find themselves
to belong to a big group of customers supporting the same business
-- a result of herd behavior. Groups of intermediate sizes near
the characteristic size $s_{0}$ will be the hardest to please.
When a group decides not to make a deal, members of the group tend
to seek alternative opinion by combining with another group and
there is a higher probability that groups with size around the
characteristic size are combined. From the business point of view,
businesses may find it easier to satisfy a small group of
customers.  Big businesses with a big group of customers enjoy the
frame and customers are self-satisfied.  Business with a
characteristic size $s_{0}$ finds it difficult to satisfy its
group of customers due to the diversity in demand.  In the merging
of businesses, it is more likely that businesses of size of the
characteristic size merge so as to become a bigger business with
possibly a better image to the customers.

The present model readily recovers previous extensions of the EZ
model. In the EZ model\cite{EZ}, the fragmentation rate and the
coagulation rate are simply taken to be $f(s_{i})=c(s_{i},
s_{j})\equiv1$.  The model leads to the size distribution $n_{s}
\sim s^{-\tau}$ in the steady state with $\tau$ taking on a {\em
robust} value of $5/2$, together with an exponential cutoff
setting in for large $s$ \cite{rodgers2}. Recently, it has been
shown that if the fragmentation rate and the coagulation rate are
taken to have a power law dependence on the size of the group(s)
involved, i.e., $f(s_{i})=s_{i}^{-\delta}$ and $c(s_{i},
s_{j})=s_{i}^{-\delta}s_{j}^{-\delta}$, respectively, the exponent
$\tau$ characterizing the distribution of group sizes becomes
model-dependent and takes on the value
$\tau=5/2-\delta$\cite{zheng1}.

Figure 1 shows the simulation results for the group size
distribution $n_{s}/n_{1}$ as a function of $s$ for a system with
a total of $N=10^{4}$ agents. The results are obtained within the
time window between $t=10^{5} \sim 10^{6}$ steps, after the system
approaches its steady state.  Each data point represents an
average taken over 32 independent runs. As scaling behavior is
more conveniently observed for small value of $a$\cite{EZ}, we
have taken $a=0.01$. Figure 1(a) shows the results for different
characteristic sizes of $s_{0}=1, 5, 20, 10^{4}$ with the
parameter $\delta=0.2$. For $s_{0}=1$, $n_{s}$ has a power-law
decay with an exponent of $\tau = 5/2-\delta = 2.3$ throughout the
whole range of $s$ since only the $s > s_{0}$ regime exists. For
$s_{0}=N=10^{4}$ corresponding to the largest possible
characteristic size in the system, scaling is also found
throughout the whole regime of $s$, but with a different value of
the exponent of $\tau = 5/2+\delta =2.7$. For intermediate values
of $s_{0}$, e.g., $s_{0}=5$ and $s_{0}=20$, two scaling regimes
corresponding to the exponents $2.7$ and $2.3$ are found for $s <
s_{0}$ and $s > s_{0}$, respectively. Figure 1(b) shows similar
results for a larger value of $\delta =0.5$ with $s_{0}=1, 5, 20,
10^{4}$. Crossover between the two scaling regimes are observed
again at $s=s_{0}$ for the intermediate cases of $s_{0} = 5$ and
$s_{0} = 20$. For $s_{0}=N$, however, large groups of customers
disappear and the scaling behavior no longer exist in the present
case. This feature can be understood qualitatively as the
coagulation rate is too small to allow large groups of customers
to be formed\cite{zheng2}.

\section{Mean field analysis}

The present model can be analyzed via a mean field
theory\cite{rodgers2,zheng1}.
Let $n_{s}(t)$ be the number of groups of agents with
size $s$ at time $t$. The master equations
that govern $n_{s}(t)$($s>1$) and $n_{1}(t)$ are
\begin{equation}
N \frac{\partial n_{s}}{\partial t} =
-af(s)sn_{s} + \frac{(1-a)}{N} \sum_{r=1}^{s-1}
c(r,s-r)rn_{r} (s-r)n_{s-r}
- \frac{2(1-a)sn_{s}}{N} \sum_{r=1}^{\infty} c(s,r)rn_{r},
\label{eq:ns time evol}
\end{equation}
and
\begin{equation}
N \frac{\partial n_{1}}{\partial t} =
a \sum_{r=2}^{\infty} f(r)r^{2}n_{r}
- \frac{2 (1-a)n_{1}}{N} \sum_{r=1}^{\infty} c(1,r)rn_{r},
\label{eq:n1 time evol}
\end{equation}
respectively.  Here $f(i)$ and $c(i,j)$ are the fragmentation and
coagulation rates defined in Eqs.(1) and Eq.(2).
The first term on the right hand side of Eq.(\ref{eq:ns time evol})
describes the fragmentation of a group of customers of size $s$.
The second term
describes the coagulation of two groups to form a group
of size $s$. The last
term is the coagulation of a group of size $s$ with another group.
In Eq.(\ref{eq:n1 time evol}),
the first term on the right hand side describes
the fragmentation of bigger groups into isolated
customers and the second term represents
the coagulation of a group of size unity with
another group. In the steady state,
$\frac{\partial n_{s}}{\partial t} = 0$.  We have
\begin{equation}
n_{s} = \frac{1-a}{Nasf(s) + 2(1-a)s\sum_{r=1}^{\infty} c(s,r)rn_{r}}
 \sum_{r=1}^{s-1} c(r,s-r)r(s-r)n_{r} n_{s-r}
\label{eq:ns steady state}
\end{equation}
for $s>1$, and
\begin{equation}
n_{1} = \frac{Na}{2(1-a) \sum_{r=1}^{\infty} c(1,r)rn_{r}}
 \sum_{r=2}^{\infty} f(r)r^{2} n_{r}.
\label{eq:n1 steady state}
\end{equation}

The steady state equations (\ref{eq:ns steady state})
and (\ref{eq:n1 steady state}) can be solved by using
the generating function technique\cite{zheng1}.
Following standard procedures\cite{zheng1,zheng2},
$n_{s}$ can be solved to give the scaling behavior in the limit
of $a \approx 0$ to get
\begin{equation}
n_{s} \sim Ns^{-(\frac{5}{2} + \delta)}
\label{eq:resulting ns, 1}
\end{equation}
for $s \leq s_{0}$, and
\begin{equation}
n_{s} \sim Ns^{-(\frac{5}{2} - \delta)}
\label{eq:resulting ns, 2}
\end{equation}
for $s > s_{0}$.
For finite value of $a$, the scaling behavior is masked by an
exponential cutoff.
Equations (\ref{eq:resulting ns, 1}) and
(\ref{eq:resulting ns, 2}) indicate that there are two scaling
behavior for the size distribution of groups of customers.
The exponents characterizing the power law decay
of the size distribution of groups of customers are
$\tau=5/2+\delta$ for $s \leq s_{0}$ and
$\tau=5/2-\delta$ for $s > s_{0}$, respectively.  These results are
in agreement with the numerical results given in Fig.1.

\section{Summary}

In summary, we have presented an extension of the EZ model for
opinion sharing and herd formation in a population of agents in
which there is a characteristic size $s_{0}$. The model can be
treated analytically via a mean field approach.  Both results from
numerical simulation and mean field theory indicate that the size
distribution of groups of agents $n_{s}$ exhibits two distinct
scaling behavior depending on $s \leq s_{0}$ or $s > s_{0}$. For
$s \leq s_{0}$, $n_{s} \sim s^{-(5/2 + \delta)}$, while for $s >
s_{0}$, $n_{s} \sim s^{-(5/2 - \delta)}$.  Here $\delta$ is a
model parameter and hence the exponents characterizing the scaling
behavior are tunable. Suitably interpreted, our model can be used
to represent the formation of groups of customers for certain
products produced by manufacturers.  This, in turn, leads to the
size distribution of businesses.  Within this context, the
characteristic size $s_{0}$ represents the size of a business for
which the customer group becomes too ``large" to be pleased but
too ``small" for the business to become a brand name. With its
tunable exponent, the model can be used to fit to empirical data
as it has been shown that markets in different countries are
characterized by non-universal exponents\cite{ramsden}. Recently,
it has also been shown empirically\cite{nagel} that the
distributions of U.S. establishment sizes and firm sizes in the
retail sector are characterized by power laws with an exponent of
$-2$. Within our model, such a behavior can be obtained for
$\delta=0.5$ and $s_{0}=1$.  The introduction of the two
parameters $\delta$ and $s_{0}$ thus makes the present model
highly flexible and hence can be used to model a wide variety of
phenomena in which herd formation and information transmission are
important.

\begin{center}
{\bf ACKNOWLEDGMENTS}
\end{center}

DFZ, GJR and PMH acknowledge financial
support from The China Scholarship Council, The Leverhulme Trust  and the
Research Grants Council of the Hong Kong SAR Government under grant
CUHK4241/01P, respectively.

\newpage \centerline{\bf FIGURE CAPTIONS}

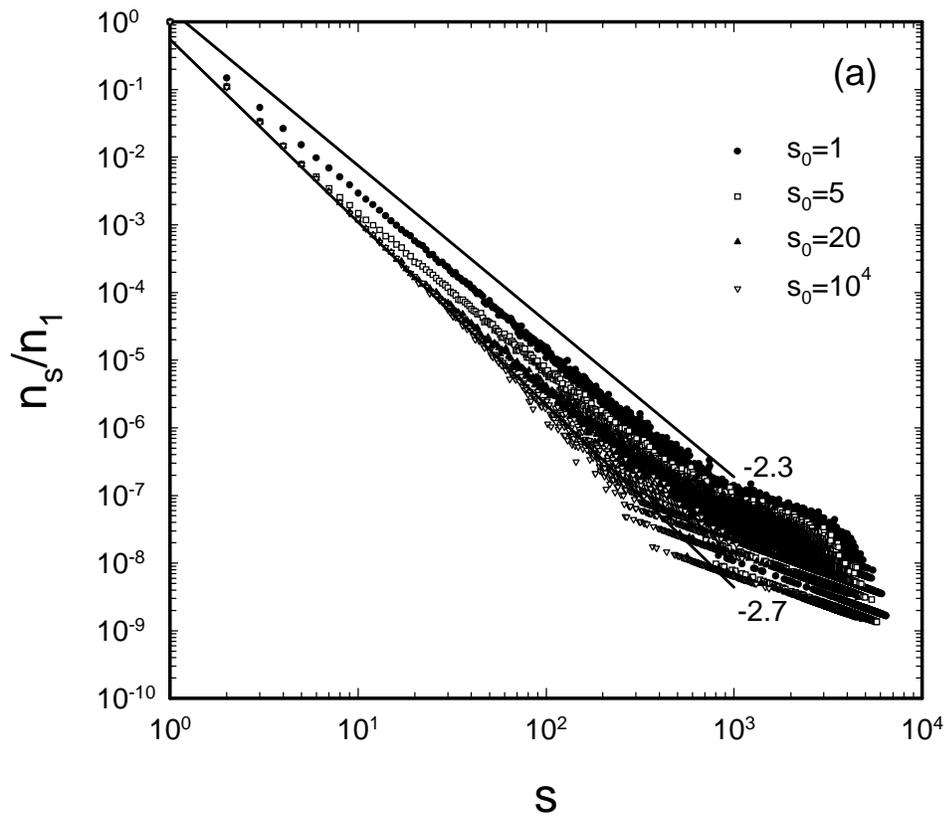
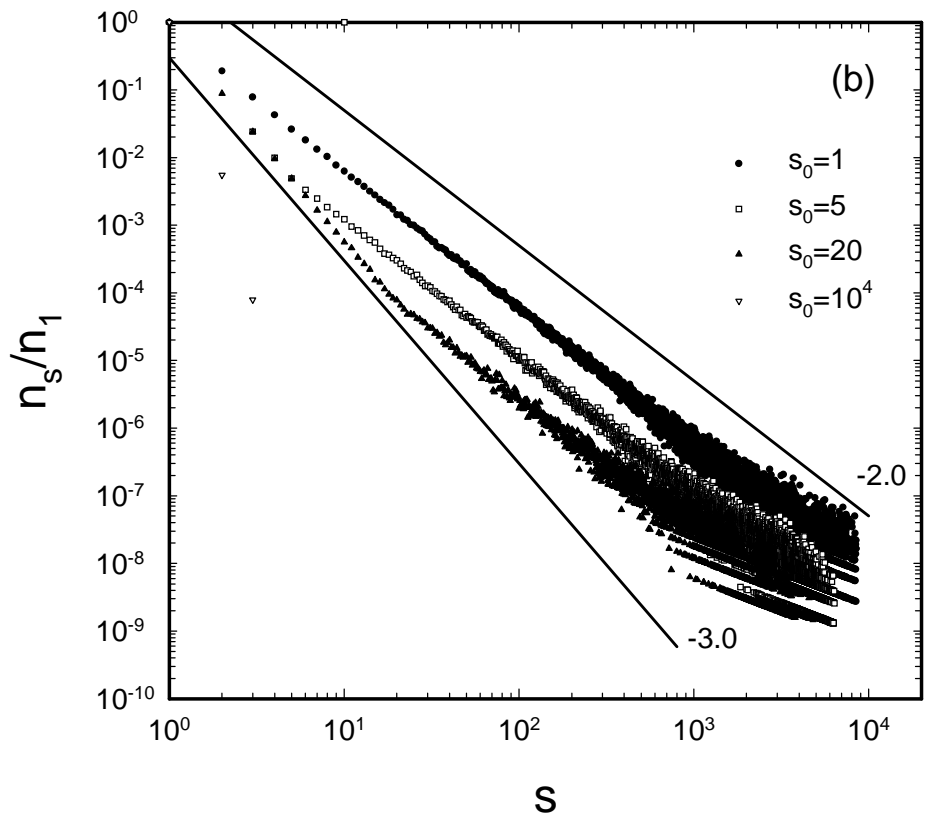
Fig. 1. The size distribution of groups of customers $n_{s}/n_{1}$
as a function of size $s$ on a log-log scale for (a)$\delta=0.2$
and (b)$\delta=0.5$. The values of the characteristic size $s_{0}$
chosen in the calculations are: $s_{0}=1, 5, 20, 10^{4}$. The
solid lines are a guide to the eye.

\end{document}